\documentclass{article}

\usepackage{arxiv}

\usepackage[english]{babel}  

\usepackage{graphicx}
\usepackage{svg}

\usepackage{tikz}
\usetikzlibrary{shapes.geometric,positioning}
\usepackage{tkz-euclide}

\usepackage{todonotes}

\usepackage{xspace}

\usepackage{amsmath}
\usepackage{amssymb}
\usepackage{mathtools}
\usepackage{multirow}

\usepackage{url}

\usepackage{cleveref}
\crefname{lstlisting}{listing}{listings}
\Crefname{lstlisting}{Listing}{Listings}
\crefname{figure}{figure}{figures}
\Crefname{figure}{Figure}{Figures}

\usepackage{paralist}

\usepackage{xcolor}
\allowdisplaybreaks

\newcommand{\cpp}{C++\xspace}

\definecolor{VividEmerald}{RGB}{0,152,116}
\colorlet{skcolor}{VividEmerald!40} % Sebastian

\definecolor{DeepSkyBlue}{RGB}{0,191,255}
\colorlet{sfcolor}{DeepSkyBlue!40} % Sara

\definecolor{Strawberry}{RGB}{252,90,141}
\colorlet{vacolor}{Strawberry!40} % Vadym

\usepackage{acro}
\newcommand{\acrodef}[2]{\DeclareAcronym{#1}{short={#1},long={#2}}}

\acrodef{SWE}{shallow water equations}
\acrodef{CFL}{Courant-Friedrichs-Lewy}
\acrodef{DG}{Discontinuous Galerkin}
\acrodef{DSL}{domain-specific language}
\acrodef{HPC}{high performance computing}
\acrodef{LSE}{linear system of equations}
\acrodef{MLUpS}{million lattice updates per second}
\acrodef{SPH}{smoothed particle hydrodynamics}
\acrodef{PDE}{partial differential equation}
\acrodef{UFL}{unified form language}

\usepackage{listings}
\usepackage{color}
\usepackage{inconsolata}

\definecolor{keywordColor}{RGB}{117,112,179}
\definecolor{commentColor}{RGB}{27,158,119}
\definecolor{stringColor}{RGB}{217,95,2}

\lstdefinelanguage{ExaSlang1}
{
  basicstyle=\ttfamily\small,
  keywords={[2]
    ApplicationHint, ApplicationHints, Cell, DiscretizationHint, DiscretizationHints, Discretize, Domain, Equation, Face\_x, Face\_y, Face\_z, Field, Knowledge, L2Hint, L2Hints, L3Hint, L3Hints, L4Hint, L4Hints, Node, Operator, Solve, SolverHint, SolverHints, \_, all, and, but, cell, coarser, coarsest, current, direction, face\_x, face\_y, face\_z, false, finer, finest, for, generate, import, in, node, not, on, order, solver, to, true, with
  },
  keywordstyle={[2]\color{keywordColor}\bfseries},
  commentstyle=\color{commentColor}, % style of comments
  stringstyle=\color{stringColor}, % style of strings
  showstringspaces=false,
  sensitive=true, % keywords are case-sensitive
  morecomment=[l]{//}, % l is for line comment
  morecomment=[s]{/*}{*/}, % s is for start and end delimiter
  morestring=[b]{'}, % defines that strings are enclosed in single quotes
  morestring=[b]{"}, % defines that strings are enclosed in double quotes
  literate={
    {\@}{@}1
  	{\\in}{$\in$}1
  	{\\partial}{$\partial$}1
  	{\\times}{$\times$}1
  	{\\Delta}{$\Delta$}1
  	{\\Omega}{$\Omega$}1
  	{\\BSin}{\textbackslash{}in}1
    {\\BSpartial}{\textbackslash{}partial}1
    {\\BStimes}{\textbackslash{}times}1
    {\\BSDelta}{\textbackslash{}Delta}1
    {\\BSOmega}{\textbackslash{}Omega}1
  },
}

\lstdefinelanguage{ExaSlang2}
{
  basicstyle=\ttfamily\small,
  keywords={[2]
    ApplicationHint, ApplicationHints, Array, Bool, Boolean, Cell, Complex, Domain, Double, Equation, Expr, Expression, Face\_x, Face\_y, Face\_z, Field, Float, Globals, Int, Integer, Knowledge, L3Hint, L3Hints, L4Hint, L4Hints, Neumann, Node, None, Operator, Real, Solve, SolverHint, SolverHints, Stencil, StencilTemplate, String, Unit, Val, Value, Var, Variable, all, and, array, bool, boolean, bottom, boundary, but, cell, coarser, coarsest, complex, current, default, double, east, equation, face\_x, face\_y, face\_z, false, finer, finest, float, for, from, generate, i0, i1, i2, import, in, int, integer, is, node, north, not, of, on, operators, prolongation, real, restriction, solver, south, store, string, times, to, top, true, unit, west, with, x, y, z
  },
  keywordstyle={[2]\color{keywordColor}\bfseries},
  commentstyle=\color{commentColor}, % style of comments
  stringstyle=\color{stringColor}, % style of strings
  showstringspaces=false,
  sensitive=true, % keywords are case-sensitive
  morecomment=[l]{//}, % l is for line comment
  morecomment=[s]{/*}{*/}, % s is for start and end delimiter
  morestring=[b]{'}, % defines that strings are enclosed in single quotes
  morestring=[b]{"}, % defines that strings are enclosed in double quotes
  literate={
    {\@}{@}1
  },
}

\lstdefinelanguage{ExaSlang3}
{
	basicstyle=\ttfamily\small,
  keywords={[2]
    ApplicationHint, ApplicationHints, Array, Bool, Boolean, Cell, Complex, Domain, Double, Equation, Expr, Expression, Face\_x, Face\_y, Face\_z, Field, Float, Func, FuncTemplate, Function, FunctionTemplate, Globals, Inst, Instantiate, Int, Integer, Knowledge, L2, L4Hint, L4Hints, Neumann, Node, None, Operator, Real, Stencil, String, Unit, Val, Value, Var, Variable, all, and, append, array, as, bc, bool, boolean, bottom, boundary, but, cell, coarser, coarsest, color, complex, count, current, default, double, east, else, face\_x, face\_y, face\_z, false, finer, finest, float, for, from, generate, i0, i1, i2, if, import, in, int, integer, jacobi, locally, loopBase, modifiers, node, north, not, of, on, or, override, prepend, prolongation, real, relax, repeat, replace, restriction, return, smootherHint, smootherStage, solve, solveFor, solver, south, string, times, to, top, true, unit, until, west, where, while, with, x, y, z
  },
  keywordstyle={[2]\color{keywordColor}\bfseries},
  commentstyle=\color{commentColor}, % style of comments
  stringstyle=\color{stringColor}, % style of strings
  showstringspaces=false,
  sensitive=true, % keywords are case-sensitive
  morecomment=[l]{//}, % l is for line comment
  morecomment=[s]{/*}{*/}, % s is for start and end delimiter
  morestring=[b]{'}, % defines that strings are enclosed in single quotes
  morestring=[b]{"}, % defines that strings are enclosed in double quotes
  literate={
    {\@}{@}1
  },
}

\lstdefinelanguage{ExaSlang4}
{
  basicstyle=\ttfamily\small,
  keywords={[2]
    Array, Bool, Boolean, CVector, Cell, ColumnVector, Complex, Domain, Edge\_Cell, Edge\_Node, Equation, Expr, Expression, Face\_x, Face\_y, Face\_z, Field, Func, FuncTemplate, Function, FunctionTemplate, Globals, Inst, Instantiate, Int, Integer, Knowledge, Layout, LayoutTransformations, Matrix, Neumann, Node, None, Operator, RVector, Real, RowVector, Set, Stencil, StencilField, StencilTemplate, String, T, Unit, Val, Value, Var, Variable, Vec2, Vec3, Vec4, Vector, active, activeSlot, advance, all, and, apply, as, bc, begin, bottom, boundary, break, but, cell, coarser, coarsest, color, communicate, communicating, communication, concat, contraction, count, current, currentSlot, default, dup, east, edge\_cell, edge\_node, else, ending, external, face\_x, face\_y, face\_z, false, finer, finest, finish, fragments, from, fromFile, ghost, i0, i1, i2, if, import, into, jacobi, locally, loop, next, nextSlot, node, noinline, north, not, of, on, only, or, over, postcomm, precomm, previous, previousSlot, prolongation, reduction, relax, rename, repeat, restriction, return, sequentially, solve, south, starting, stepping, steps, times, to, top, transform, true, until, west, where, while, with, x, y, z
  },
  keywordstyle={[2]\color{keywordColor}\bfseries},
  commentstyle=\color{commentColor}, % style of comments
  stringstyle=\color{stringColor}, % style of strings
  showstringspaces=false,
  sensitive=true, % keywords are case-sensitive
  morecomment=[l]{//}, % l is for line comment
  morecomment=[s]{/*}{*/}, % s is for start and end delimiter
  morestring=[b]{'}, % defines that strings are enclosed in single quotes
  morestring=[b]{"}, % defines that strings are enclosed in double quotes
  literate={
    {\@}{@}1
  },
}

\lstdefinelanguage{Config}
{
  basicstyle=\ttfamily\small,
  keywords={[2]
    import
    },
  keywordstyle={[2]\color{keywordColor}\bfseries},
  commentstyle=\color{commentColor}, % style of comments
  stringstyle=\color{stringColor}, % style of strings
  showstringspaces=false,
  sensitive=true, % keywords are case-sensitive
  morecomment=[l]{//}, % l is for line comment
  morecomment=[s]{/*}{*/}, % s is for start and end delimiter
  morestring=[b]{'}, % defines that strings are enclosed in single quotes
  morestring=[b]{"}, % defines that strings are enclosed in double quotes
}

\begin{document}

\title{Towards whole program generation of quadrature-free discontinuous Galerkin methods for the shallow water equations}

\author{  Sara Faghih-Naini \\
  Department of Computer Science\\
  Friedrich-Alexander-Universit{\"a}t \\
  Erlangen-N{\"u}rnberg, Germany\\
  Alfred Wegener Institute\\
   Helmholtz Centre for Polar and Marine Research\\
   Bremerhaven, Germany
   \And
   Sebastian Kuckuk \\
	Department of Computer Science\\
  Friedrich-Alexander-Universit{\"a}t \\
  Erlangen-N{\"u}rnberg, Germany\\
   \And
   Vadym Aizinger \\
Department of Mathematics\\
  Friedrich-Alexander-Universit{\"a}t \\
  Erlangen-N{\"u}rnberg, Germany\\
  Alfred Wegener Institute\\
   Helmholtz Centre for Polar and Marine Research\\
   Bremerhaven, Germany \\
   \And
   Daniel Zint \\
	Department of Computer Science\\
  Friedrich-Alexander-Universit{\"a}t \\
  Erlangen-N{\"u}rnberg, Germany\\
     \And
   Roberto Grosso \\
	Department of Computer Science\\
  Friedrich-Alexander-Universit{\"a}t \\
  Erlangen-N{\"u}rnberg, Germany\\
     \And
   Harald K{\"o}stler \\
	Department of Computer Science\\
  Friedrich-Alexander-Universit{\"a}t \\
  Erlangen-N{\"u}rnberg, Germany\\
  \texttt{harald.koestler@fau.de}\\
}

\maketitle              % typeset the header of the contribution

\begin{abstract}
The shallow water equations (SWE) are a commonly used model to study tsunamis, tides, and coastal ocean circulation. However, there exist various approaches to discretize and 
solve them efficiently. Which of them is best for a certain scenario is often not known and, in addition, depends heavily on the used HPC platform. From a~simulation 
software perspective, this places a~premium on the ability to adapt easily to different numerical methods and hardware architectures. One solution to this problem is to apply code
generation techniques and to express methods and specific hardware-dependent implementations on different levels of abstraction. This allows for a separation of concerns and makes 
it possible, e.g., to exchange the discretization scheme without having to rewrite all low-level optimized routines manually. 
In this paper, we show how code for an advanced quadrature-free discontinuous Galerkin (DG) discretized shallow water equation solver can be generated. Here, we follow the multi-layered 
approach from the ExaStencils project that starts from the continuous problem formulation, moves to the discrete scheme, spells out the numerical algorithms, and, finally,
maps to a representation that can be transformed to a distributed memory parallel implementation by our in-house Scala-based source-to-source compiler.
Our contributions include: A~new quadrature-free discontinuous Galerkin formulation, an~extension of the class of supported computational grids, and an~extension of 
our toolchain allowing to evaluate discrete integrals stemming from the DG discretization implemented in Python.   
As first results we present the whole toolchain and also demonstrate the convergence of our method for higher order DG discretizations.
\end{abstract}

\section{Introduction}

ExaStencils\footnote{\url{http://www.exastencils.org/}} provides a multi-layered domain specific language to model various simulation applications in computational science and engineering that involve the solution of partial differential equations (PDEs). It does not rely on other simulation software packages to solve them iteratively but generates the whole simulation program from input data, application parameters, and problem descriptions formulated in an external domain specific language. 

In this work, we consider the shallow water equations (SWEs), the main type of model used for prediction of floods caused by tsunami and storm surges as well as for many other problems in oceanography, meteorology, and coastal engineering. 
We have already shown scalability of a~generated basic SWE solver on a~GPU cluster~\cite{kuckuk2018whole}. Now we extend our previous work in several directions:
\begin{compactenum}
\setlength\itemsep{0pt}
\item apply a~state-of-the-art discontinuous Galerkin (DG) method based discretization
that we reformulate to become quadrature-free
\item provide a~Python interface for our external domain specific language enabling the user to start with a~symbolic algebra representation of a~discrete problem and transform it automatically to a~C++ code via our ExaStencils toolchain
\item add support for more general computational grids necessary for computing more realistic scenarios  
\end{compactenum}

\paragraph{Quadrature-free DG discretization:}
In a~quadrature-free DG scheme, all element and face integrals in the discrete formulation are evaluated analytically instead of using quadrature rules. 
The advantage of a~quadrature-free vs. a~quadrature integration is the fact that the former eliminates the innermost loop over quadrature points offering thereby a~better code optimization potential. 
This approach is not new (see, e.g., \cite{Atkins1998,Lockard1999}); however, until now, it has mostly been applied to linear problems or problems with product-type nonlinearities (e.g., advection terms in Euler equations) that only involve integrals of polynomials easily evaluated analytically. By exploiting one the main strengths of the local DG (LDG) method, the mixed re-formulation of the original PDEs, we mo\-di\-fy the SWE system making it amenable to the quadrature-free methodology in spite of the presence of fraction-type nonlinearities. Whereas, in the original LDG framework~\cite{CockburnShu1998}, such mixed formulations are employed to replace higher-order derivatives by lower-order ones, we took this technique one step further in~\cite{BungertAF2017,AizingerBF2018} to handle nonlinearities in the PDE for mean-curvature flow. A~similar idea is used in the current work to deal with the fraction terms in the SWE. 

Quadrature-free DG formulations are an~alternative to the tensor-product DG formulations. Schemes based on the latter approach evaluate multidimensional integrals as products of one-dimensional ones achieving in this way a~higher computational intensity particularly for high-order discretizations. While very efficient for element shapes naturally represented in a~tensor-product fashion (quads, hexahedra, etc.), this approach is less popular for simplicial elements (triangles, tets), although such generalization also exist (e.g., Dubiner bases~\cite{Dubiner1991}).

%In oceanography typically fluid flows are considered, where vertical scales of motion are much smaller than horizontal ones.
%Based on this assumption, the general Euler equations can be specialized to the \acp{SWE}~\cite{whitham2011linear}.
%For them, several different approaches for discretization and various implementations exist.
Other discretization examples for the SWE include a~recent DG~\cite{wintermeyer2017entropy} implementation, finite differences~\cite{casulli1990semi,asai2016coupled} or generalized finite differences~\cite{li2017generalized}, or even a~mesh-free \ac{SPH} approach~\cite{bankole2015semi}.
Efficient implementations are found in literature~\cite{wittmann2017high,poppl2016swe} using, e.g., a~quadtree data structure on GPU~\cite{vacondio2017non}, Sierpinski curves~\cite{bader2010dynamically}, and dynamically adaptive spacetree grids~\cite{weinzierl2014block}.
Some implementations are also running on distributed memory systems~\cite{lai2016parallel}. 

\paragraph{Domain-specific languages for HPC:}
In addition to software libraries, \acp{DSL} also became increasingly prominent in computational science and engineering applications in recent years.
A~more detailed overview of popular approaches is found in~\cite{SKHTL18procieee}.
Particularly relevant ones are Firedrake, STELLA, Mint and SPIRAL.
Firedrake~\cite{Rathgeber2016} is an automated toolchain for solving PDEs specified in a~\ac{DSL} embedded in Python.
It employs the \ac{UFL}~\cite{UFL} and the FEniCS form compiler of the FEniCS project~\cite{fenicsbook}.
STELLA~\cite{GOFBS15} targets stencil codes on structured grids and uses, in contrast to ExaStencils, an internal \ac{DSL} embedded into \cpp that is based on template meta programming.
Mint~\cite{unat2011mint} is a~programming model for GPUs and focuses on the source-to-source compilation of annotated C to CUDA.
Finally, SPIRAL~\cite{Pueschel2011} provides abstractions for linear transforms and other mathematical functions.

Our approach builds upon the ExaStencils code generator that relies on an~external multi-layered \ac{DSL} called ExaSlang to map from different levels of abstraction: First, the continuous problem description; next, the discretization; then the numerical algorithm specification; and, finally, a~C-like representation of the whole simulation code. The ExaStencils toolchain has been now extended by adding support for a~symbolic algebra math-like specification of the discrete problem within Python. The latter is then mapped automatically to the C-like ExaSlang layer, from where a~regular code generation process goes on.    

\paragraph{Expanding support for more general computational grids:}
Initially, ExaStencils only supported regular grids with sparse matrix-vector multiplications easily described by stencil operations. In a~step-by-step fashion, support for more general computational grids including staggered and grids with locally varying mesh size has been added.  
In order to be able to run simulations for realistic ocean domain geometries that cannot be accurately approximated by structured grids, our final goal is to support block-structured grids obtained from a~coarse unstructured grid refined regularly. To this end, a~grid generator is being currently developed to create grids with a~block-structured topology as required by the Exastencils framework with first results for complex domains presented in~\cite{Zint2018}.
% While grids for simple domains can be already generated directly within ExaStencils, for complex domains, the grid generator is used as an~external tool, and the generated grids are read in at the start of the simulation. 

The paper is structured as follows: In Sec.~\ref{sec:swe} the continuous problem, its DG discretization, and the algebraic representation of element integrals are described followed by the mapping to executable code via Python and the ExaStencils compiler.
Numerical verification of the generated code is presented in Sec.~\ref{sec:num}.

\section{A shallow water equations generator}
\label{sec:swe}

\begin{figure}[th!]
\centering
    \includegraphics[width=0.9\textwidth]{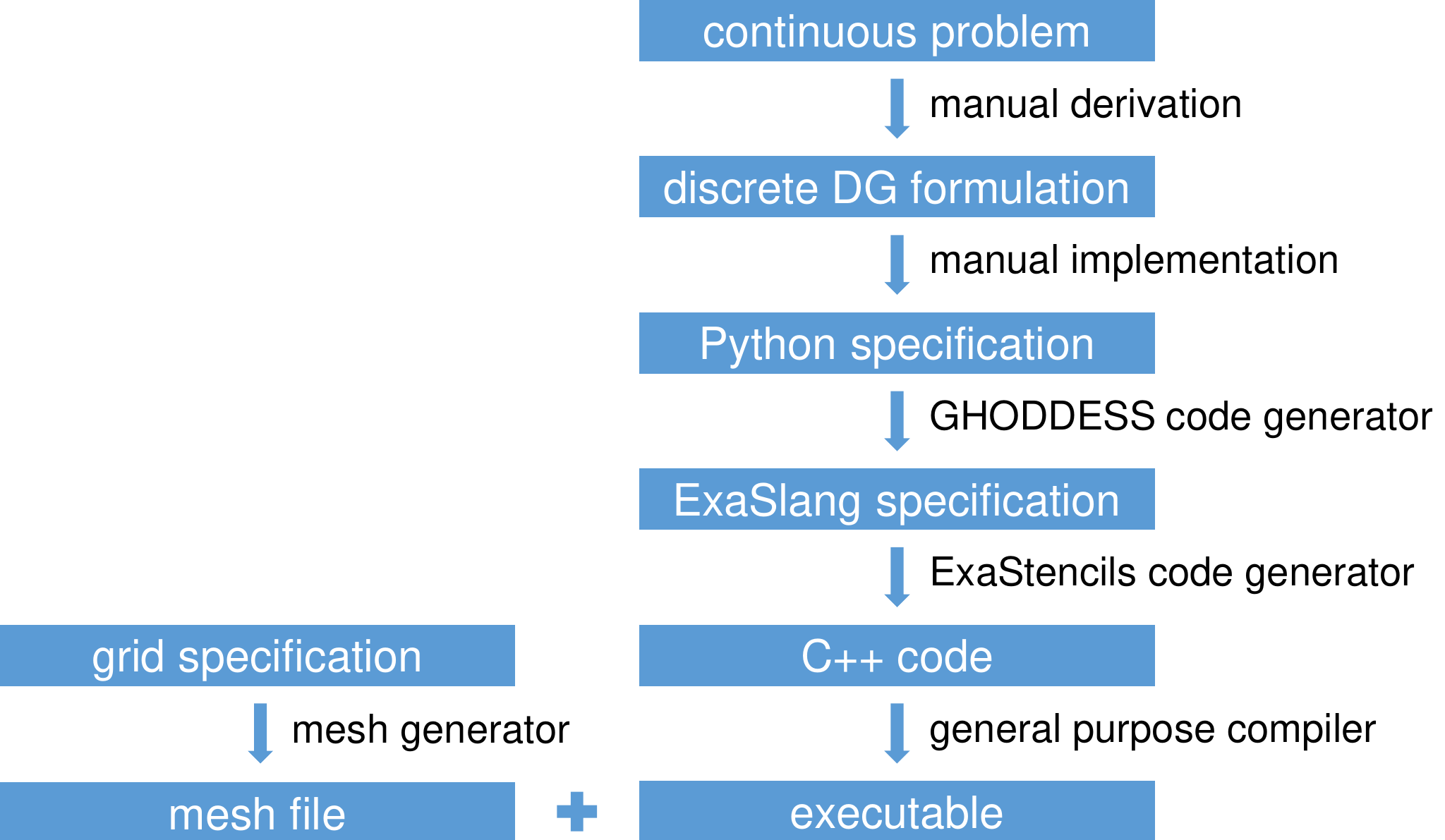}
  \caption{Extended ExaStencils toolchain for generating code for DG discretized SWEs.}
  \label{fig:ghoddessexascheme}
\end{figure}

The main steps of our approach to map a~continuous mathematical problem description to an efficient simulation code are sketched in Fig.~\ref{fig:ghoddessexascheme}. Next, we describe each of these steps in more detail.

\subsection{Continuous formulation}
The classical 2D SWE are obtained from the vertically integrated Navier-Stokes equations under the additional assumptions of a hydrostatic pressure and a vertically uniform horizontal velocity:
\begin{align}
\label{mass}
&\partial_t \xi+\nabla \cdot \boldsymbol{q}=0,\\
\label{momentum}
&\partial_t \boldsymbol{q}
+\nabla \cdot \left(\boldsymbol{qq}^T /H \right)
+\tau_{bf}\boldsymbol{q}
+\left( \begin{smallmatrix} 0 & -f_c\\ f_c & 0 \end{smallmatrix} \right) \boldsymbol{q}
+gH\nabla\xi
=\boldsymbol{F}. 
\end{align}
They are defined on some 2D domain $\Omega$, and $\xi$ represents the elevation of the free water surface with respect to some datum (e.g., the mean sea level). By $H = h_b + \xi$, we denote the total fluid depth with $h_b$ representing the bathymetric depth, $\boldsymbol{q} \equiv (U,V)^T$ is the depth integrated horizontal velocity field, $f_c$ is the Coriolis coefficient, $g$ is the gravitational acceleration, and $\tau_{bf}$ is the bottom friction coefficient. Effects of variable atmospheric pressure, and tidal potentials are expressed through the body force $\boldsymbol{F}$.\\
Defining $\boldsymbol{c} \coloneqq (\xi, U,V)^T$ system~\eqref{mass}--\eqref{momentum} is given in the following compact form:
\begin{align}
\partial_t \boldsymbol{c}+ \nabla \cdot \boldsymbol{A}= \boldsymbol{r}(\boldsymbol{c}),
\label{compact}
\end{align}
where 
\begin{align}
\label{eq:def_A}
&\boldsymbol{A}=\begin{pmatrix}
U & V\\
\frac{U^2}{H}+\frac{1}{2}g(H^2-h_b^2)&\frac{U V}{H}\\
\frac{U V}{H}&\frac{V^2}{H}+\frac{1}{2}g(H^2-h_b^2)
\end{pmatrix}\\ 
&\text{and } \nonumber\\
\label{eq:def_r}
&\boldsymbol{r}(\boldsymbol{c})=\begin{pmatrix}
0\\
-\tau_{bf}U+f_cV+g\xi \partial_x h_b+F_x\\
-\tau_{bf}V-f_cU+g\xi \partial_y h_b+F_y\\
\end{pmatrix}.
\end{align}
In the remainder of this paper, the boundary conditions are assumed to be of Dirichlet type; 
however, a~full set of boundary conditions needed for realistic simulations will be included in our implementation in the near future.

\subsection{Discretization using a~quadrature-free DG method}
\label{sec:dis}

Our discretization of the SWE system~\eqref{compact} is generally based on the scheme realized in our UTBEST solver~\cite{DawsonAizinger2002a,aiz02}. But, instead of a~standard quadrature-based evaluation of integrals, we utilize a~quadrature-free DG formulation enhanced to deal with nonlinearities in form of fractions. 
For this purpose, we introduce the depth averaged velocity $\boldsymbol{u}=(u,v)^T$ given by $ \boldsymbol{q} = \boldsymbol{u}H$ and extend system~\eqref{compact} by equations for $u$ and $v$:
%a new variable $\tilde{\boldsymbol{u}}=(u,v)^T$is introduced.
%With $\boldsymbol{\tilde{c}}\coloneqq(\xi, \boldsymbol{\tilde{u}}^T)^T\eqqcolon (\xi, \tilde{U}, \tilde{V})^T$, equation (\ref{compact}) becomes: 
%Then 
%our system \eqref{compact} is written as follows:
\begin{align}
\label{compact_qf_1}
\partial_t \boldsymbol{c}+ \nabla \cdot \boldsymbol{\tilde{A}}= \boldsymbol{r}(\boldsymbol{c}),\\
\label{compact_qf_2}
\boldsymbol{u} H = \boldsymbol{q},
\end{align}
where $H^2-h_b^2=\xi^2+2h_b\xi+h_b^2-h_b^2=\xi(H+h_b)$ and thus 
\begin{align}
\boldsymbol{\tilde{A}}=\begin{pmatrix}
U & V\\
Uu+\frac{1}{2}g\xi(H+h_b)&U v\\
V u& V v+\frac{1}{2}g\xi(H+h_b)
\end{pmatrix}.
\label{def_A_h_qf}
\end{align}

Let $\{\mathcal{T}_\Delta\}_{\Delta>0}$ be a family of triangulations  of $\Omega\subset\mathbb{R}^2$, and let $\Omega_e, e\in \{0,\dots,E\}$ be elements of $\mathcal{T}_\Delta$. We obtain the variational formulation of system \eqref{compact_qf_1}--\eqref{compact_qf_2} by multiplication with sufficiently smooth test functions $\boldsymbol{\phi}$ and $\boldsymbol{\psi}$ and integration by parts  on each element $\Omega_e \in \mathcal{T}_\Delta$, which yields:
\begin{align}
&\left(\partial_t,\boldsymbol{\phi}\right)_{\Omega_e}-\left(\boldsymbol{\tilde{A}}, \nabla \boldsymbol{\phi}\right)_{\Omega_e}+\langle\boldsymbol{\tilde{A}}\cdot\boldsymbol{n}, \nabla \boldsymbol{\phi}\rangle_{\partial\Omega_e}= \left(\boldsymbol{r}(\boldsymbol{c}),\boldsymbol{\phi}\right)_{\Omega_e},\\
&\left(\boldsymbol{u} H, \boldsymbol{\psi} \right)_{\Omega_e}
= \left(\boldsymbol{q}, \boldsymbol{\psi} \right)_{\Omega_e},
\end{align}
where $(\cdot,\cdot)_{\Omega_e}$ and $\langle\cdot,\cdot\rangle_{\partial \Omega_e}$ represent the $L^2-$ scalar products on elements and edges, respectively.

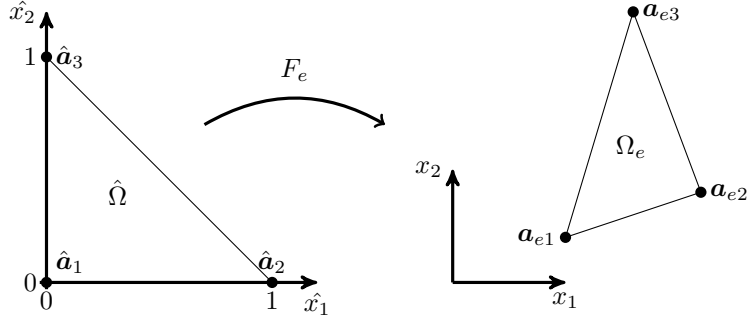
\begin{figure}
\centering
\begin{tikzpicture}[
    scale=3,
    axis/.style={very thick, ->, >=stealth' },
    important line/.style={thick},
    every node/.style={color=black},
    ]
    % axis left coordinate system
    \draw[axis] (-0.03,0)  -- (1.2,0) node(xline)[below]
        {$\hat{x_1}$};
    \draw[axis] (0,-0.03) -- (0,1.2) node(yline)[left] {$\hat{x_2}$};
     % unit triangle
	\draw (0,0) node[anchor=south west]{$\hat{\boldsymbol{a}}_1$}
  	-- (1,0) node[anchor=south]{$\hat{\boldsymbol{a}}_2$}
  	-- (0,1) node[anchor=west]{$\hat{\boldsymbol{a}}_3$}
  	-- cycle;
  \tkzDefPoint(0,0){A} 
  \tkzDefPoint(1,0){B} 
  \tkzDefPoint(0,1){C} 
  \tkzLabelPoint[left](A){$0$}
  \tkzLabelPoint[below](A){$0$}
  \tkzLabelPoint[below](B){$1$}
  \tkzLabelPoint[left](C){$1$}
  \tkzDefPoint(0.4,0.4){D} 
  \tkzLabelPoint[left](D){$\hat{\Omega}$}
  % bend arrow
  \tkzDefPoint(0.7,0.7){E}
  \tkzDefPoint(1.5,0.7){F}
  \tkzDefPoint(1.1,0.85){G}
  \tkzLabelPoint[above](G){$F_e$}
  \draw [->, very thick] (E) to [out=30,in=150] (F);
  % axis right coordinate system
  \draw[axis] (1.8,0)  -- (2.3,0) node(xline)[below]
        {$x_1$};
  \draw[axis] (1.8,0) -- (1.8,0.5) node(yline)[left] {$x_2$};
  % physical triangle
  \draw (2.3,0.2) node[anchor= east]{$\boldsymbol{a}_{e1}$}
  -- (2.9,0.4) node[anchor=west]{$\boldsymbol{a}_{e2}$}
  -- (2.6,1.2) node[anchor=west]{$\boldsymbol{a}_{e3}$}
  -- cycle;
  \tkzDefPoint(2.3,0.2){H} 
  \tkzDefPoint(2.9,0.4){I} 
  \tkzDefPoint(2.6,1.2){J}
  \tkzDefPoint(2.7,0.6){K} 
  \tkzLabelPoint[left](K){$\Omega_e$}
  \foreach \n in {A,B,C,H,I,J}
  \node at (\n)[circle,fill,inner sep=1.5pt]{};
\end{tikzpicture}
\caption{Affine-linear mapping $F_e$ from the reference triangle $\hat{\Omega} = \left\{\hat{\boldsymbol{a}}_1, \hat{\boldsymbol{a}}_2, \hat{\boldsymbol{a}}_2 \right\} = \left\{[0,0]^T, [1,0]^T, [0,1]^T \right\}$ to physical triangle $\Omega_e = \left\{\boldsymbol{a}_{e,1}, \, \boldsymbol{a}_{e,2}, \boldsymbol{a}_{e,3} \right\}$.}
\label{mapping}
\end{figure}

\noindent
Now denoting by $\mathbb{P}^k(\Omega_e)$ the polynomial spaces of order $k$ on $\Omega_e$, we obtain the discrete formulation using test functions $\boldsymbol{\phi}_\Delta \in \mathbb{P}^k(\Omega_e)^3, \, \boldsymbol{\psi}_\Delta \in \mathbb{P}^k(\Omega_e)^2$:
\begin{align}
\label{variational_disc_1}
\begin{split}
&\left(\partial_t \boldsymbol{c}_\Delta,\boldsymbol{\phi}_\Delta\right)_{\Omega_e}
-\left(\boldsymbol{\tilde{A}}(\boldsymbol{c}_\Delta, \boldsymbol{u}_\Delta), \nabla \boldsymbol{\phi}_\Delta\right)_{\Omega_e}
+\langle\boldsymbol{\hat{\tilde{A}}}(\boldsymbol{c}_\Delta, \boldsymbol{u}_\Delta,\boldsymbol{c}^+_\Delta, \boldsymbol{u}^+_\Delta, \boldsymbol{n}),\boldsymbol{\phi}_\Delta \rangle_{\partial\Omega_e}\\[0pt]
& \qquad = \left(\boldsymbol{r}(\boldsymbol{c}_\Delta, \boldsymbol{u}_\Delta),\boldsymbol{\phi}_\Delta\right)_{\Omega_e},
\end{split}\\
\label{variational_disc_2}
&\left(\boldsymbol{u}_\Delta H_\Delta, \boldsymbol{\psi}_\Delta \right)_{\Omega_e}
= \left(\boldsymbol{q}_\Delta, \boldsymbol{\psi}_\Delta \right)_{\Omega_e},
\end{align}
where $\boldsymbol{n}$ is a~unit normal to $\partial \Omega_e$, and $\boldsymbol{\tilde{A}}\cdot\boldsymbol{n}$ is approximated on $\partial\Omega_e$ by a~numerical flux $\boldsymbol{\hat{\tilde{A}}}(\boldsymbol{c}_\Delta, \boldsymbol{u}_\Delta,\boldsymbol{c}^+_\Delta, \boldsymbol{u}^+_\Delta, \boldsymbol{n})$ (here, we utilize the Lax-Friedrichs flux~\cite{HajdukHAR2018}) that depends on discontinuous values of the solution on element $\Omega_e$ (without superscript) and on its edge neighbors (superscript~$^+$).
$\boldsymbol{c}_\Delta$ and $\boldsymbol{u}_\Delta$ are the DG approximations to $\boldsymbol{c}$ and $\boldsymbol{u}$ and can be represented as
\begin{align}
\boldsymbol{c}_\Delta(t,\boldsymbol{x})|_{\Omega_e} &=(\xi_\Delta, U_\Delta, V_\Delta)^T(t,\boldsymbol{x})=\sum_{j=1}^3\sum_{i=1}^{K(k)} c_{ei}^j\varphi_{ei}(\boldsymbol{x})\,\boldsymbol{e}_j, \label{c_h}\\
 \boldsymbol{u}_\Delta(t,\boldsymbol{x}) |_{\Omega_e} &=(u_\Delta, v_\Delta)^T(t,\boldsymbol{x})=\sum_{j=1}^2\sum_{i=1}^{K(k)} u_{ei}^j\varphi_{ei}(\boldsymbol{x})\,\boldsymbol{e}_j\label{c_h_tilde}
\end{align}
with $\boldsymbol{e}_j$ denoting the $j$-th unit vector in $\mathbb{R}^3$ in~\eqref{c_h} or $\mathbb{R}^2$ in~\eqref{c_h_tilde}. A~basis of space $\mathbb{P}^k(\Omega_e)$ consisting of $\varphi_{ei}(\boldsymbol{x}), \,i=1, \ldots, K(k)$ is defined using a~mapping from the corresponding reference basis
\begin{align*}
\varphi_{ei}(\boldsymbol{x})=\begin{cases}\hat{\varphi}_i\left(\boldsymbol{F}_e^{-1}(\boldsymbol{x})\right) & x\in \Omega_e,\\0& \text{otherwise},\end{cases}\quad i \in \{1,\dots,K(k)\}
\end{align*}
where $F_e: \hat{\Omega}\rightarrow\Omega_e,\; \hat{\boldsymbol{x}}\rightarrow \boldsymbol{x}\coloneqq \boldsymbol{B}_e\hat{\boldsymbol{x}}+\boldsymbol{a}_{e1}$ is the affine-linear transformation (see Fig.~\ref{mapping}) from the reference triangle onto triangle $\Omega_e$ with
\begin{align*}
\boldsymbol{B}_e = \begin{bmatrix} B^e_{1,1} & B^e_{1,2} \\
B^e_{2,1} &B^e_{2,2}\end{bmatrix}\coloneqq \begin{bmatrix} \boldsymbol{a}_{e2}-\boldsymbol{a}_{e1} & \boldsymbol{a}_{e3}-\boldsymbol{a}_{e1}\end{bmatrix}
\end{align*}
and $\boldsymbol{a}_{e1}, \boldsymbol{a}_{e2}, \boldsymbol{a}_{e3}$ denoting the vertex coordinates of $\Omega_e$.

The number of basis functions $K(k)$ is dependent on the respective polynomial space and has the following values: $K(0)\!=\!1, \, K(1)\!=\!3, \, K(2)\!=\!6$, and $K(3)\!=\!10$.
The basis functions on the reference triangle $\hat{\Omega}$ employed in our implementation are given by:
\small
\begin{equation*}
\left.
\begin{array}{l}
\left.\!\!
\begin{array}{l}
\left.\!\!
\begin{array}{l}
\hat{\varphi}_1(\hat{\vec{x}})=\sqrt{2}, \left. \;\; \right\} \boldsymbol{\mathbb{P}^0}\\
\hat{\varphi}_2(\hat{\vec{x}})= 2 - 6\hat{x}_1,\\
\hat{\varphi}_3(\hat{\vec{x}})= \sqrt{12} (1 - \hat{x}_1 - 2\hat{x}_2),
\end{array} \right\} \; \boldsymbol{\mathbb{P}^1}\\
\hspace{0.1cm}\hat{\varphi}_4(\hat{\vec{x}})= \sqrt{6}\left(1-8\hat{x}_1+10\hat{x}_1^2\right),\\
\hspace{0.1cm}\hat{\varphi}_5(\hat{\vec{x}})= 
\sqrt{3}\left(-1-4\hat{x}_1+5\hat{x}_1^2+12\hat{x}_2-15\hat{x}_2^2\right),\\
\hspace{0.1cm}\hat{\varphi}_6(\hat{\vec{x}})=
\sqrt{45}\left(1-4\hat{x}_1+3\hat{x}_1^2-4\hat{x}_2+8\hat{x}_1\hat{x}_2+3\hat{x}_2^2\right),
\end{array} \right\} \; \boldsymbol{\mathbb{P}^2}\\
\hspace{0.2cm}\hat{\varphi}_7(\hat{\vec{x}})= \sqrt{8}\left(-1+15\hat{x}_1-45\hat{x}_1^2+35\hat{x}_1^3\right),\\
\hspace{0.2cm}\hat{\varphi}_8(\hat{\vec{x}})= \sqrt{24}\left(-1+13\hat{x}_1-33\hat{x}_1^2+21\hat{x}_1^3+2\hat{x}_2-24\hat{x}_1\hat{x}_2+42\hat{x}_1^2\hat{x}_2\right),\\
\hspace{0.2cm}\hat{\varphi}_9(\hat{\vec{x}})= \sqrt{40}\left(-1+9\hat{x}_1-15\hat{x}_1^2+7\hat{x}_1^3+6\hat{x}_2-48\hat{x}_1\hat{x}_2+42\hat{x}_1^2\hat{x}_2-6\hat{x}_2^2+42\hat{x}_1\hat{x}_2^2\right),\\
\hspace{0.05cm}\hat{\varphi}_{10}(\hat{\vec{x}})= \sqrt{56}\left(-1+3\hat{x}_1 -3\hat{x}_1^2+\hat{x}_1^3+12\hat{x}_2 -24\hat{x}_1\hat{x}_2+12\hat{x}_1^2\hat{x}_2-30\hat{x}_2^2+30\hat{x}_1\hat{x}_2^2+20\hat{x}_2^3\right).
\end{array} \right\} \; \boldsymbol{\mathbb{P}^3}
\end{equation*}
\normalsize

% \begin{align*}
% \mathbb{P}^2(\hat{\Omega}) \left\{ 
% \begin{array}{rcl}
% \begin{array}{c}
% \\\mathbb{P}^1(\hat{\Omega})\\ \\ \end{array} & 
% \left\{\begin{array}{r}\mathbb{P}^0(\hat{\Omega})\; \Big\{\\ \\ \\ \end{array} \right.
% &\begin{array}{rcl}
% \hat{\varphi}_1(\hat{\vec{x}})&=&\sqrt{2}\,,\\
% \hat{\varphi}_2(\hat{\vec{x}})&=& 2 - 6\hat{x}_1\,,\\
% \hat{\varphi}_3(\hat{\vec{x}})&=& 2 \sqrt{3} (1 - \hat{x}_1 - 2\hat{x}_2)\,,
% \end{array}\\
% &&\begin{array}{rcl}
% \hat{\varphi}_4(\hat{\vec{x}})&=& \sqrt{6}\big((10\hat{x}_1-8)\hat{x}_1+1\big)\,,\\
% \hat{\varphi}_5(\hat{\vec{x}})&=& 
% \sqrt{3}\big((5\hat{x}_1-4)\hat{x}_1+(-15\hat{x}_2+12)\hat{x}_2-1\big)\,,\\
% \hat{\varphi}_6(\hat{\vec{x}})&=& 
% 3\sqrt{5}\big((3\hat{x}_1+8\hat{x}_2-4)\hat{x}_1+(3\hat{x}_2-4)\hat{x}_2+1\big)\,.
% \end{array}
% \end{array}\right.
% \end{align*}

\paragraph{Algebraic representation of element integrals:}
For compactness, we show an~algebraic representation of our discrete scheme only for element integrals. 
Inserting the basis representations \eqref{c_h}, \eqref{c_h_tilde} into system \eqref{variational_disc_1}, \eqref{variational_disc_2} and testing the first equation with $\boldsymbol{\phi}_\Delta = \varphi_{ep} \boldsymbol{e}_1$ we obtain for $p \in  {1,\dots,K(k)}$:
\begin{align}
&\left(\boldsymbol{\tilde{A}}(\boldsymbol{c}_\Delta, \boldsymbol{u}_\Delta), \nabla (\varphi_{ep} \boldsymbol{e}_1) \right)_{\Omega_e}
= \sum_{l=1}^2 \sum_{i=1}^{K(k)} c_{ei}^{l+1}\int_{\Omega_e} \frac{\partial\varphi_{ep}} {\partial x_l}\varphi_{ei}\mathrm{d}x\nonumber\\
&\quad =\sum_{i=1}^{K(k)} \Bigg{[}\frac{c_{ei}^2}{|\mathrm{det}(\boldsymbol{B}_e)|}\left({B}^e_{2,2}\int_{\hat{\Omega}}  \frac{\partial\hat{\varphi}_{p}}{\partial \hat{x}_1} \hat{\varphi}_{i}\mathrm{d}\hat{x} - {B}^e_{2,1} \int_{\hat{\Omega}}  \frac{\partial\hat{\varphi}_{p}}{\partial \hat{x}_2} \hat{\varphi}_{i} \mathrm{d}\hat{x}\right)\nonumber\\
&\qquad +\frac{c_{ei}^3}{|\mathrm{det}(\boldsymbol{B}_e)|}\left(-{B}^e_{1,2}\int_{\hat{\Omega}}  \frac{\partial\hat{\varphi}_{p}}{\partial \hat{x}_1} \hat{\varphi}_{i}\mathrm{d}\hat{x} + {B}^e_{1,1} \int_{\hat{\Omega}}  \frac{\partial\hat{\varphi}_{p}}{\partial \hat{x}_2} \hat{\varphi}_{i}\mathrm{d}\hat{x}\right)\Bigg{]}, 
\label{mass_basis}
\intertext{
For $\boldsymbol{\phi}_\Delta = \varphi_{ep} \boldsymbol{e}_2$  and $p \in  {1,\dots,K(k)}$:}
& \left(\boldsymbol{\tilde{A}}(\boldsymbol{c}_\Delta, \boldsymbol{u}_\Delta), \nabla (\varphi_{ep} \boldsymbol{e}_2) \right)_{\Omega_e} 
=\sum_{j=1}^2\left[\sum_{i,m=1}^{K(k)} c_{ei}^2u_{em}^{j}\int_{\Omega_e}  \frac{\partial\varphi_{ep}}{\partial_{x_j}}\varphi_{ei}\varphi_{em}\mathrm{d}x\right]\nonumber\\
&\qquad+\sum_{i,m=1}^{K(k)}\frac{g}{2} c_{ei}^{1}c_{em}^{1}\int_{\Omega_e}  \frac{\partial\varphi_{ep}}{\partial_{x_1}}\varphi_{ei}\varphi_{em}\mathrm{d}x+\sum_{i=1}^{K(k)} gh_bc_{ei}^{1}\int_{\Omega_e}  \frac{\partial\varphi_{ep}}{\partial_{x_1}} \varphi_{ei} \mathrm{d}x\nonumber\\
&\quad=\sum_{i,m=1}^{K(k)} \Bigg{[}\frac{c_{ei}^2u_{em}^{1}}{|\mathrm{det}(\boldsymbol{B}_e)|^{3/2}}\left({B}^e_{2,2}\int_{\hat{\Omega}}  \frac{\partial\hat{\varphi}_{p}}{\partial \hat{x}_1} \hat{\varphi}_{i}\hat{\varphi}_{m}\mathrm{d}\hat{x} - {B}^e_{2,1} \int_{\hat{\Omega}}  \frac{\partial\hat{\varphi}_{p}}{\partial \hat{x}_2} \hat{\varphi}_{i}\hat{\varphi}_{m} \mathrm{d}\hat{x}\right)\nonumber\\
&\qquad+\frac{c_{ei}^2u_{em}^{2}}{|\mathrm{det}(\boldsymbol{B}_e)|^{3/2}}\left(-{B}^e_{1,2}\int_{\hat{\Omega}}  \frac{\partial\hat{\varphi}_{p}}{\partial \hat{x}_1} \hat{\varphi}_{i}\hat{\varphi}_{m}\mathrm{d}\hat{x} + {B}^e_{1,1} \int_{\hat{\Omega}}  \frac{\partial\hat{\varphi}_{p}}{\partial \hat{x}_2} \hat{\varphi}_{i}\hat{\varphi}_{m} \mathrm{d}\hat{x}\right)\Bigg{]}\nonumber\\
&\qquad+\sum_{i,m=1}^{K(k)}\frac{g}{2 |\mathrm{det}(\boldsymbol{B}_e)|^{3/2}} c_{ei}^{1}c_{em}^{1}\left({B}^e_{2,2}\int_{\hat{\Omega}}  \frac{\partial\hat{\varphi}_{p}}{\partial \hat{x}_1} \hat{\varphi}_{i}\hat{\varphi}_{m}\mathrm{d}\hat{x} - {B}^e_{2,1} \int_{\hat{\Omega}}  \frac{\partial\hat{\varphi}_{p}}{\partial \hat{x}_2} \hat{\varphi}_{i}\hat{\varphi}_{m} \mathrm{d}\hat{x}\right)\nonumber\\
&\qquad+\sum_{i=1}^{K(k)} \frac{gh_bc_{ei}^{1}}{|\mathrm{det}(\boldsymbol{B}_e)|} \left({B}^e_{2,2}\int_{\hat{\Omega}}  \frac{\partial\hat{\varphi}_{p}}{\partial \hat{x}_1} \hat{\varphi}_{i}\mathrm{d}\hat{x} - {B}^e_{2,1} \int_{\hat{\Omega}}  \frac{\partial\hat{\varphi}_{p}}{\partial \hat{x}_2} \hat{\varphi}_{i} \mathrm{d}\hat{x}\right)
\label{momentum_basis_u}
\end{align}
with a similar expression for $\boldsymbol{\phi}_\Delta = \varphi_{ep} \boldsymbol{e}_3$ and rather trivial representations for element integrals arising from~\eqref{variational_disc_2}.
%\todo{VA: The term with the time derivative should be removed}

%\section{Mapping to Code}

\subsection{Mapping Math to Python}
\label{sec:map}

Since the quadrature-free formulation detailed above is a~new element of our toolchain, it is a~natural starting point for code generation. An abstract representation for the terms should be, one the one hand, as close as possible to the mathematical expression and, on the other, easy to transform to an~efficient code.
Another requirement is that the format or used environment of the abstract representation is accepted by its users. 
To meet all these requirements we have chosen to use sympy\footnote{\url{https://www.sympy.org/}} within Python as the starting point. Python is one of the most often used languages currently, and sympy is a symbolic algebra package in Python with a~rich functionality.  
Our new Python module to transform the DG scheme from a symbolic sympy expression into ExaSlang is called
GHODDESS (Generation of Higher-Order Discretizations Deployed as ExaSlang Specifications). 
To generate code for the quadrature-free DG scheme, basic abstractions for, e.g., the basis functions have been implemented in GHODDESS just as classes representing triangles and data fields.  
Currently, we support quadrilateral grids only, where each element is divided into two differently oriented triangles in order to obtain a triangular grid. These special grids are obtained from our grid generator.
The algebraic representations of the discrete scheme shown in Sec.~\ref{sec:dis} for element, edge, and boundary integrals are formulated in sympy. 
As an example consider the first part of element integral for $\xi$ from \eqref{mass_basis}:
\begin{align}
\sum_{i=1}^{K(k)} \frac{c_{ei}^2}{|\mathrm{det}(\boldsymbol{B}_e)|}\left({B}^e_{2,2}\int_{\hat{\Omega}}  \frac{\partial\hat{\varphi}_{p}}{\partial \hat{x}_1} \hat{\varphi}_{i}\mathrm{d}\hat{x} - {B}^e_{2,1} \int_{\hat{\Omega}}  \frac{\partial\hat{\varphi}_{p}}{\partial \hat{x}_2} \hat{\varphi}_{i} \mathrm{d}\hat{x}\right)
\end{align}
that is translated to
\small
\begin{lstlisting}[language=Python,label=lst:python_TODO,caption={Part of element integral in Python representation.}]
sum(cu(tri.orientation, k) / tri.detB * (
    tri.B[1, 1] * sp.integrate(sp.Derivative(basisFcts[p], x, 1) 
    * basisFcts[k], (y, 0, 1 - x), (x, 0, 1))
  - tri.B[1, 0] * sp.integrate(sp.Derivative(basisFcts[p], y, 1) 
    * basisFcts[k] , (y, 0, 1 - x), (x, 0, 1))) 
  for k in range(d))
\end{lstlisting}
\normalsize

Information about the triangles such as the orientation or the determinant of the mapping from the reference triangle are stored in an object $\textit{tri}$.
Sympy is then used to evaluate integrals analytically which also allows us to perform symbolic algebra transforms on the integrals.

\subsection{Mapping Python to ExaSlang}

To allow a mapping from the symbolic representation to ExaSlang, sympy expressions are enriched with a~few required abstractions such as field symbols that correspond to accesses to ExaSlang fields that store quantities defined on the computational domain.  
GHODDESS then maps to an auxiliary knowledge file holding parameters that guide the generation process as well as ExaSlang specifications on layer 3 and 4.
The bulk part is emitted to layer 4 where the setup and solver phase are implemented. 
More information on the ExaSlang concept and its layers can be found in~\cite{schmitt2014exaslang,kuckuk2018whole}.
For higher orders DG approximations, the layer 4 file can easily grow to several MB in size and take more than one hour to generate.
The main reason for that is the quadrature-free scheme which involves symbolic integral evaluations and the complete unrolling of all loops, e.g., over basis functions.
With increasing order of the DG method, the size of the expressions increases cubically in the number of local basis functions.
Thus, one easily ends up with several million nodes in the abstract syntax tree (AST) resulting in a~noticeable slow-down of the code transforms. 

%in OceanModeling/ghoddess/swe/updates.py: 

\subsection{Mapping ExaSlang to Code}
\label{sec:exa2code}

The ExaStencils code generator is then capable of parsing the ExaSlang code, applying transforms for certain low-level optimizations on the resulting AST and pretty-printing to C++ or CUDA code combined with MPI for distributed memory architectures. 
Currently, both the ExaStencils code generator and the C++ compiler, e.g., gcc take more than one hour for higher order DG discretizations due to the large size of the expressions -- as described before.
The overall runtime of a~simulation can be in a similar range depending on the size of the computational grid and the order of the discretization.
However, larger grids do not require more time for the code generation.
Strategies to speed up the overall workflow are a~part of our ongoing work.

\section{Numerical results}
\label{sec:num}

Our implementation of the shallow water equations is planned to include an~integrated generator for block-structured meshes. 
Since this functionality is still in a~standalone testing stage, the grids employed for the runs in the current section are generated externally and read at runtime.
% The grids were designed with a block structured topology as required by the Exastencils framework, but they do not include realistic ocean domain geometries yet. 
% While grids for simple domains can be already generated directly, more sophisticated methods are necessary for complex domains. 
% An example for block structured grid generation on complex domains is given by Zint et al. \cite{Zint2018}.

%\todo{HK: grid paragraph is repetion from introduction, better here??}

The main goal of the following numerical studies is to verify the implementation and to 
demonstrate the performance of our quadrature-free formulation for a~range of benchmarks with analytically specified solutions.

%\subsection{Grid generation}

\subsection{Convergence test on a randomly perturbed regular mesh}

\begin{figure}[h!]
\centering
    \includegraphics[width=0.32\textwidth, trim=20 20 40 50, clip]{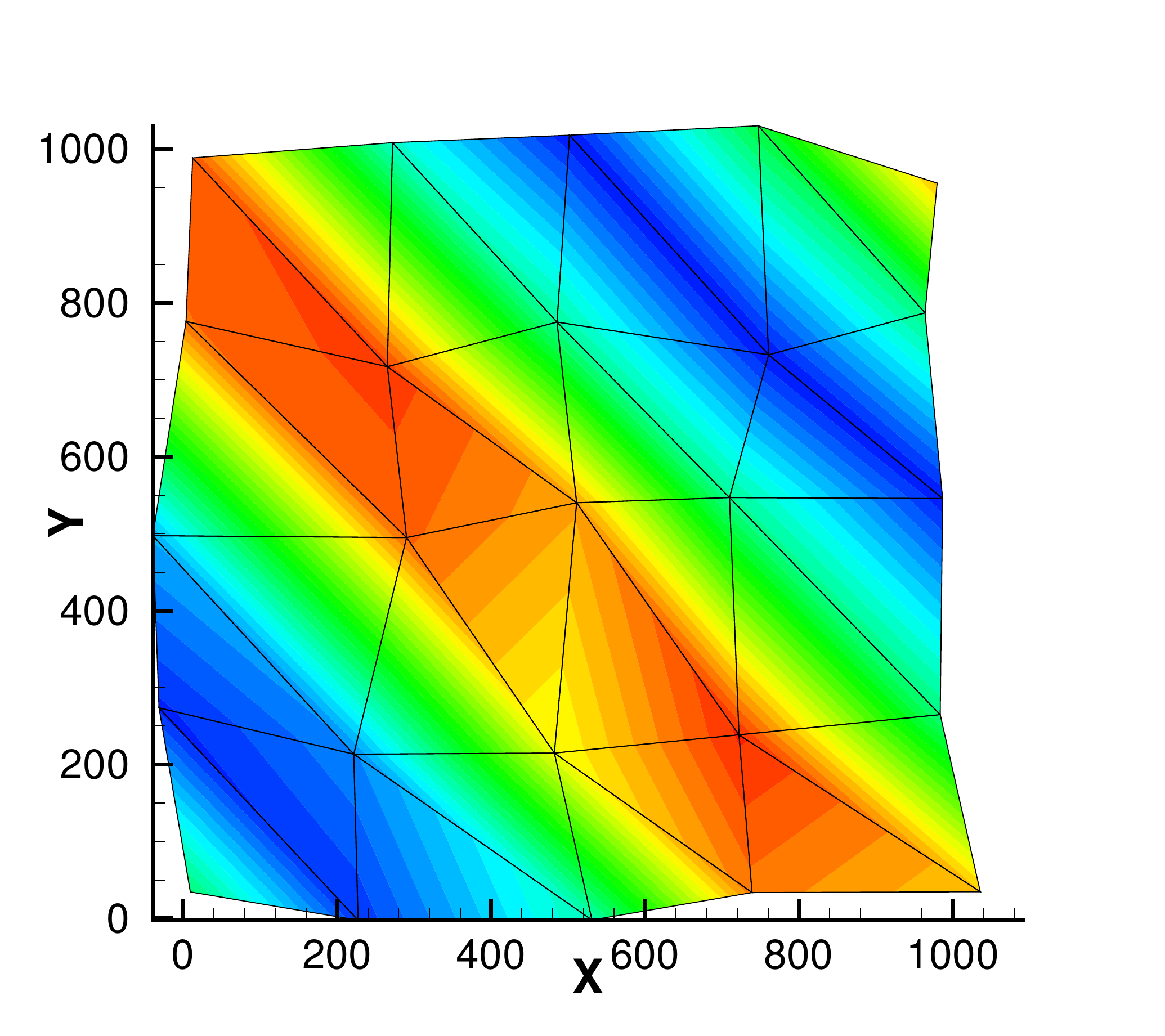}
    \includegraphics[width=0.32\textwidth, trim=20 20 40 50, clip]{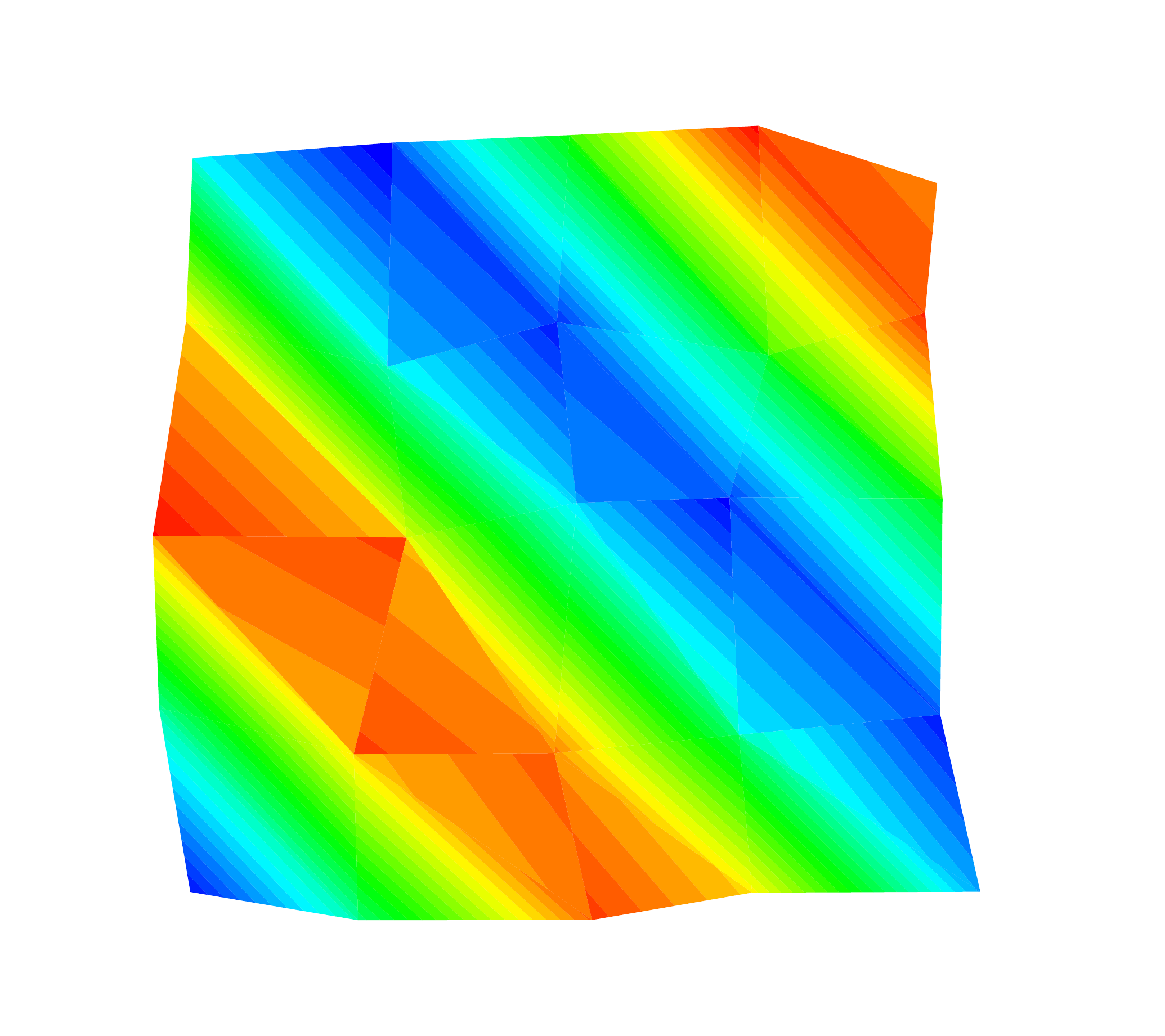}
    \includegraphics[width=0.32\textwidth, trim=20 20 40 50, clip]{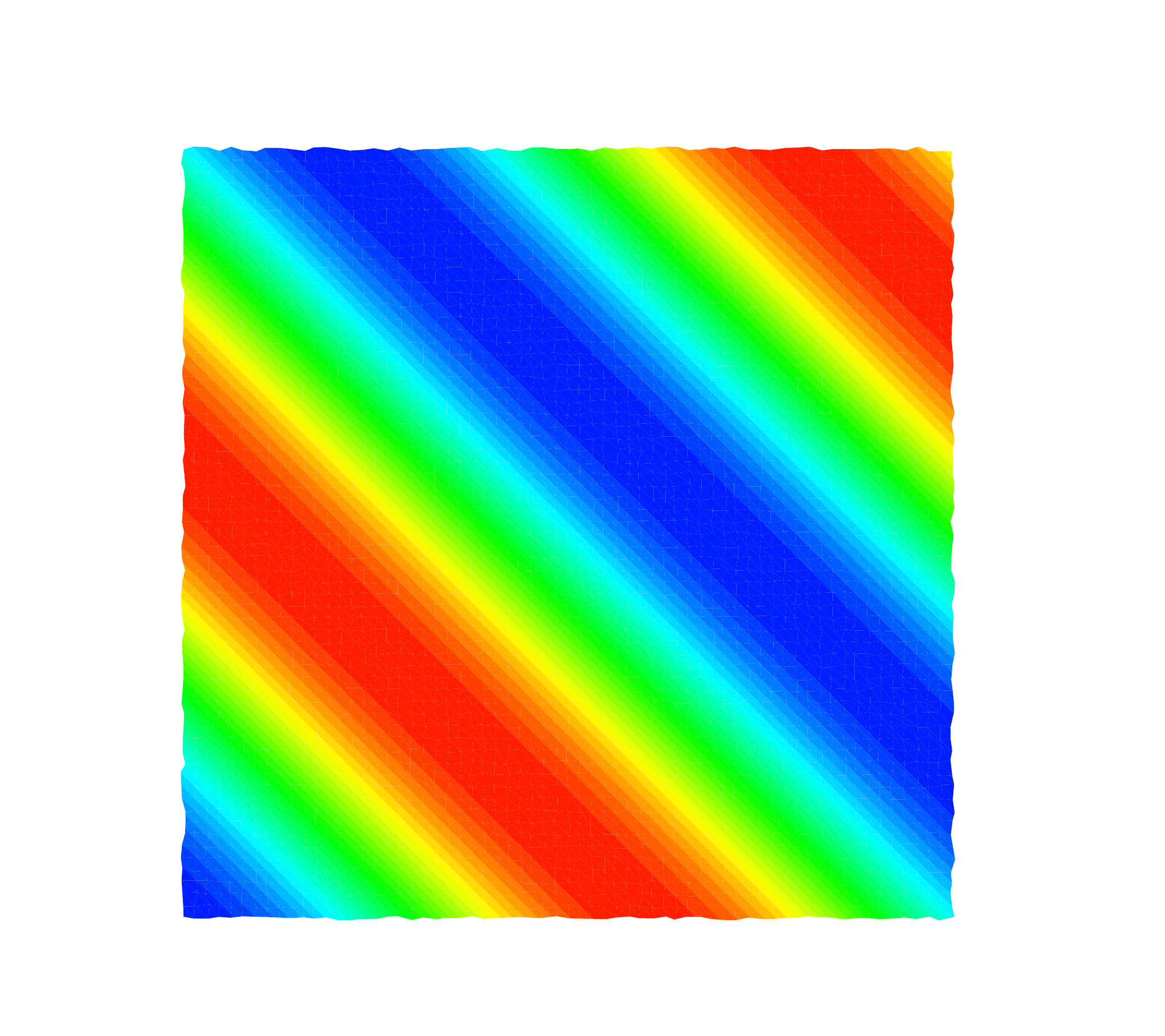}\\[2pt]
\includegraphics[width=0.8\textwidth, trim=0 0 0 480, clip]{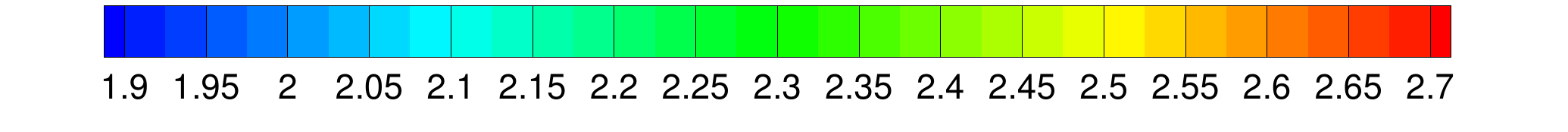}
  \caption{Square domain with perturbed mesh and piecewise linear (k=1) DG discretization: Level-2 mesh and the initial condition (left), final solution on the level-2 mesh (middle), final solution on the level-6 mesh (right).}
  \label{fig:random}
\end{figure}

For the first test, a~rectangular domain $\Omega = [0\pm r,1000\pm r]\times [0\pm r,1000\pm r]$ is used, where $r$ is a~randomly generated perturbation of up to $20\%$ of the edge length (cf. Fig.~\ref{fig:random}). The artificially manufactured analytical solution is given by
\begin{align*}
\xi(\boldsymbol{x},t)&=2+y_a-2 \,C_a\, \sin\left(\frac {\pi(x_0+x_1+C_t\,t)}{600}\right),\\
U(\boldsymbol{x},t)&=2\,y_a+C_a\,C_t\sin\left(\frac {\pi(x_0+x_1+C_t\,t)}{600}\right),\\
V(\boldsymbol{x},t)&=y_a+C_a\,C_t\sin\left(\frac {\pi(x_0+x_1+C_t\,t)}{600}\right)
\end{align*}
with the bathymetry specified as
\[
h_b =1+\frac{1}{1000} x_0+\frac{2}{1000}x_1
\]
and suitable initial and Dirichlet boundary conditions. Simulations were run for $t \in(0,1500)$ with the time step $\Delta t =0.5$ chosen small enough to make the time discretization errors negligible compared to those of the spatial discretization.
The remaining parameters are chosen as follows:  $C_a=0.2,\, C_t=0.2,\, y_a=0.3$.

\begin{figure}[h!]
  \begin{center}
    \includegraphics[width=0.9\textwidth, trim=40 15 40 70, clip]{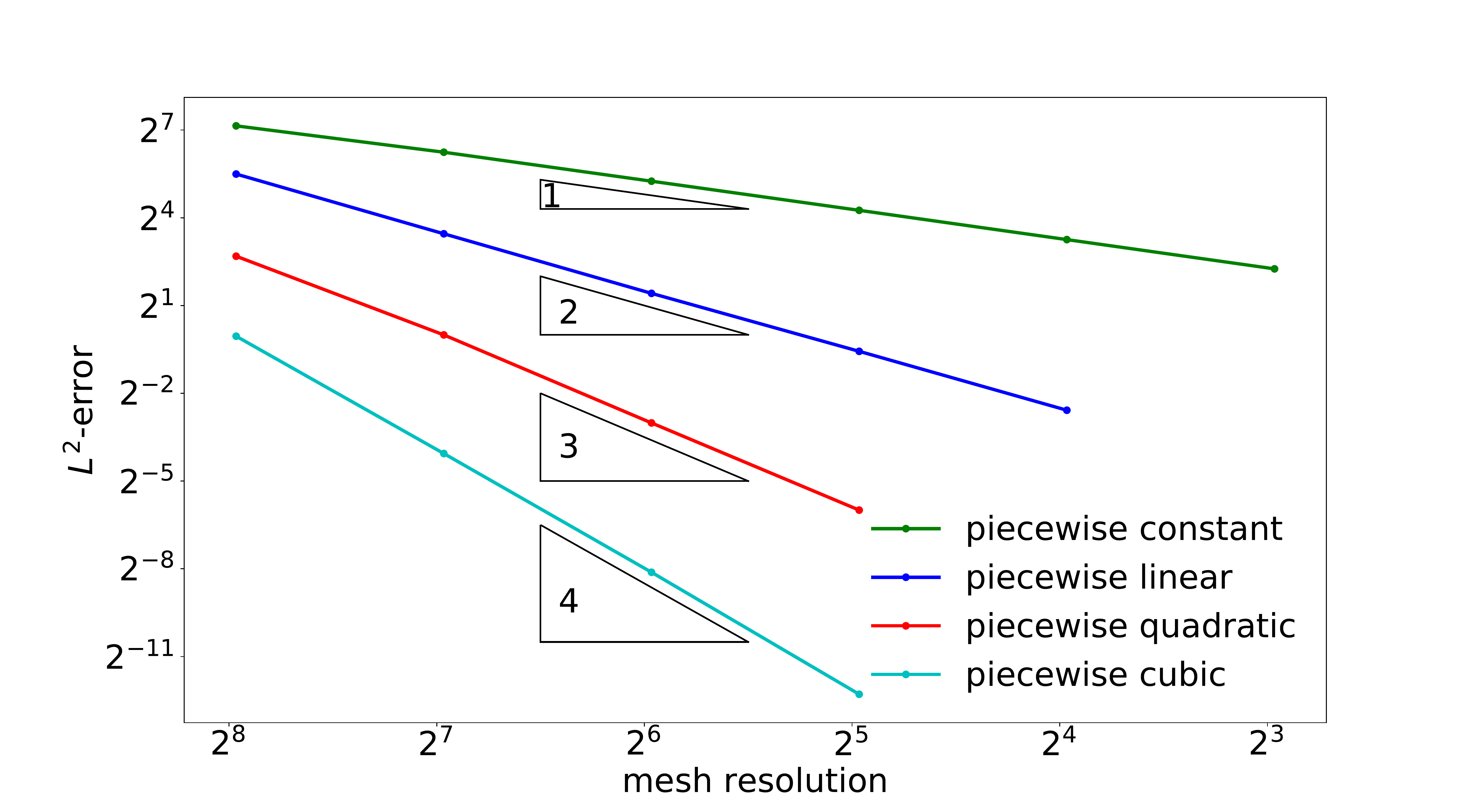}
  \end{center}
  \caption{$L^2$-errors for $\xi$ vs. the mesh resolution (cell width of the unperturbed mesh). }
  \label{fig:convergence}
\end{figure}

Fig.~\ref{fig:random} illustrates the mesh and the initial condition (left) and shows the final solution on the coarsest (level-2, middle) and finest (level-6, right) meshes for the piecewise linear (k=1) DG discretization. The errors for DG discretization spaces ranging from piecewise constants (k=0) to piecewise cubics (k=3) and the corresponding experimental convergence rates are listed in Tab.~\ref{tab:convergence} and plotted in Fig.~\ref{fig:convergence}. The expected convergence rates are demonstrated for all primary unknowns (surface elevation and depth integrated velocity components) and the results are consistent with those presented in~\cite{Aizinger2004}. Due to increasing computational (and code generation) costs, runs for higher order DG discretizations stop at coarser mesh resolutions than the low order runs; nevertheless, the $L^2$-errors for higher order approximations are much lower.

\begin{table}[h!]
\setlength{\tabcolsep}{3pt}
\centering
%d=1\\
\begin{tabular}{cllclclc}
\hline
DG order (k)  & $\#$ elements & Err($\xi$) & EOC($\xi$) & Err($U)$ & EOC($U$)& Err($V$) & EOC($V$)\\
\hline
\multirow{6}{*}{0} 
&32	&	141.24  &-	& 88.049	& -	& 96.451	& -\\ 		
&128	&	75.619	& 0.90	& 62.621	& 0.49	&	64.906	& 0.57\\ 
&512	&	38.070	& 0.99	& 35.016	& 0.84	&	35.753	& 0.86\\ 
&2048	&	19.093	& 1.00	& 18.458	& 0.92	&	18.953	& 0.92\\ 
&8192	&	9.5399	& 1.00	& 9.4733	& 0.96	& 9.7749	& 0.96\\ 
&32768	&	4.7698	& 1.00	& 4.7945	& 0.98	& 4.9692	& 0.98\\ 
\hline
\multirow{6}{*}{1} 
&32   & 45.085 & -    & 123.68 & -    & 101.33 & -    \\ 
&128  & 10.949 & 2.04 & 20.386  & 2.60 & 19.396  & 2.39 \\
&515  & 2.6693  & 2.04 & 6.2844   & 1.70 & 5.4303   & 1.84 \\
&2048 & 0.6760  & 1.98 & 1.2573   & 2.32 & 1.2092   & 2.17 \\
&8192 & 0.1674  & 2.01 & 0.2651   & 2.25 & 0.2411   & 2.33 \\ 
\hline
\multirow{4}{*}{2} 
&32   & 6.4361 & -    & 14.090 & -    & 21.492 & -    \\ 
&128	 & 0.9973 & 2.69 & 2.1901  & 2.69 & 2.5439  & 3.08 \\ 
&512  & 0.1239 & 3.01 & 0.2579  & 3.09 & 0.2419  & 3.39 \\ 
&2048 & 0.0157 & 2.98 & 0.0263  & 3.29 & 0.0225  & 3.43 \\ 
\hline
\multirow{4}{*}{3} 
&32   & 0.9669 & -    & 1.8941 & -    & 2.6533 & -    \\ 
&128	 & 0.0608 & 3.99 & 0.0985 & 4.27 & 0.1167 & 4.51 \\ 
&512	 & 0.0036 & 4.08 & 0.0077 & 3.68 & 0.0069 & 4.07 \\ 
&2048 & 0.0002 & 3.88 & 0.0004 & 4.15 & 0.0004 & 4.23 \\
\hline 
  \end{tabular}
\caption{$L^2$-errors Err($\cdot$) and experimental orders of convergence EOC($\cdot$) for the square domain with perturbed mesh and DG discretization orders $k=0, 1, 2, 3$.}
\label{tab:convergence}
\end{table}

%The number of basis functions is $d = { 1, 3, 6, 10 }$ for orders ${0,1,2,3}$.   

% \begin{compactitem}
%  \item $d == { 1, 3, 6 }$, optionally $10$ if ready in time
%  \item grid size increases until error is below $5 \cdot 10^{-order}$ or similar
%  \item grids with 20 \% max displacement in each dimension
%  \item with lambda
%  \item $t == 0.5$
%  \item until $t_{max} = 1500$
% \end{compactitem}

%Current results are available under \url{https://docs.google.com/spreadsheets/d/1s6f4IM-r7z5F5nkmVmR61adKiJaEsT349u8S6t0kR9o/edit?usp=sharing}, in the table \texttt{paper\_results}

\subsection{Circular domain}

\begin{figure}[h!]
\centering
    \includegraphics[width=0.32\textwidth, trim=20 20 40 50, clip]{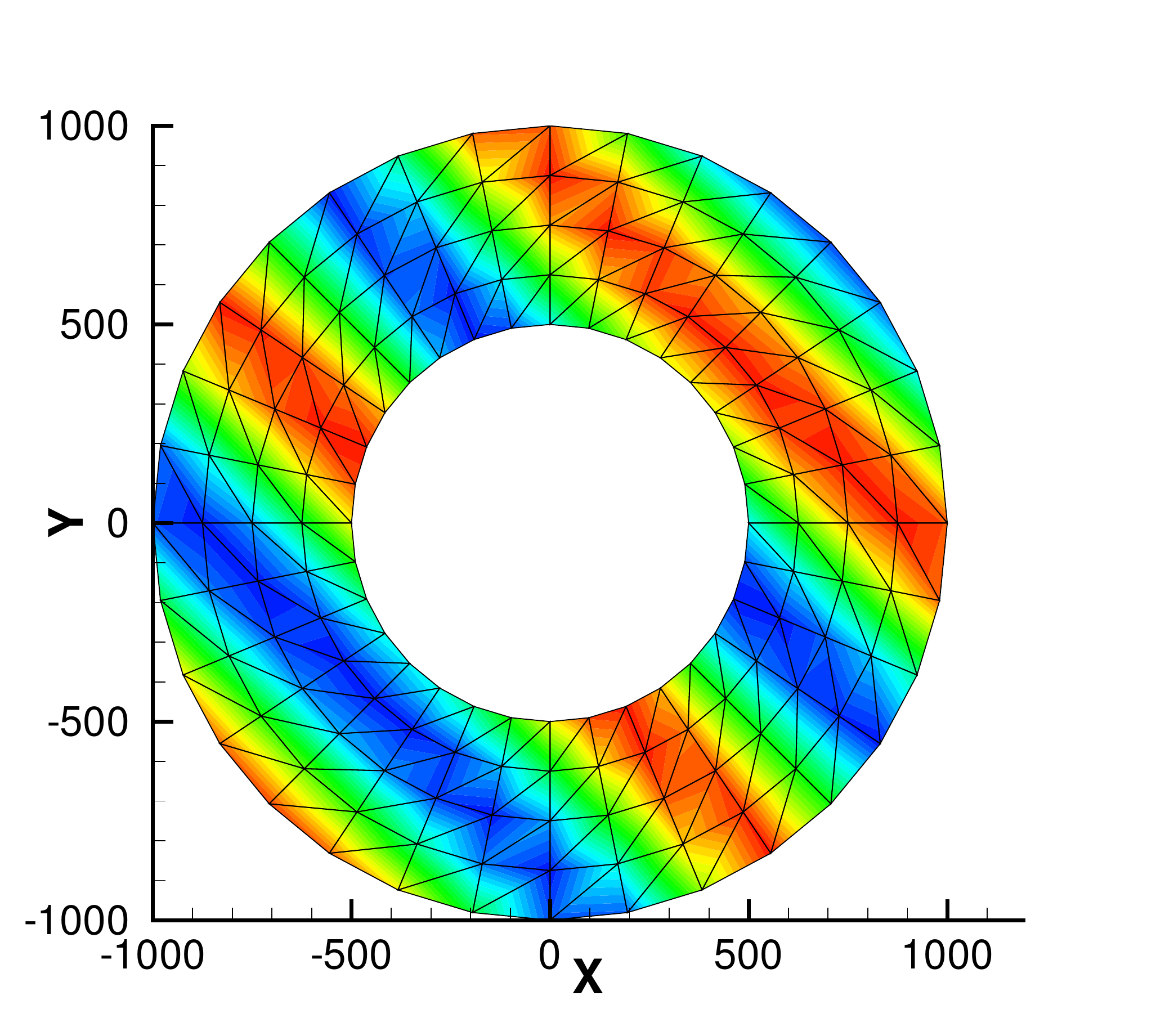}
    \includegraphics[width=0.32\textwidth, trim=20 20 40 50, clip]{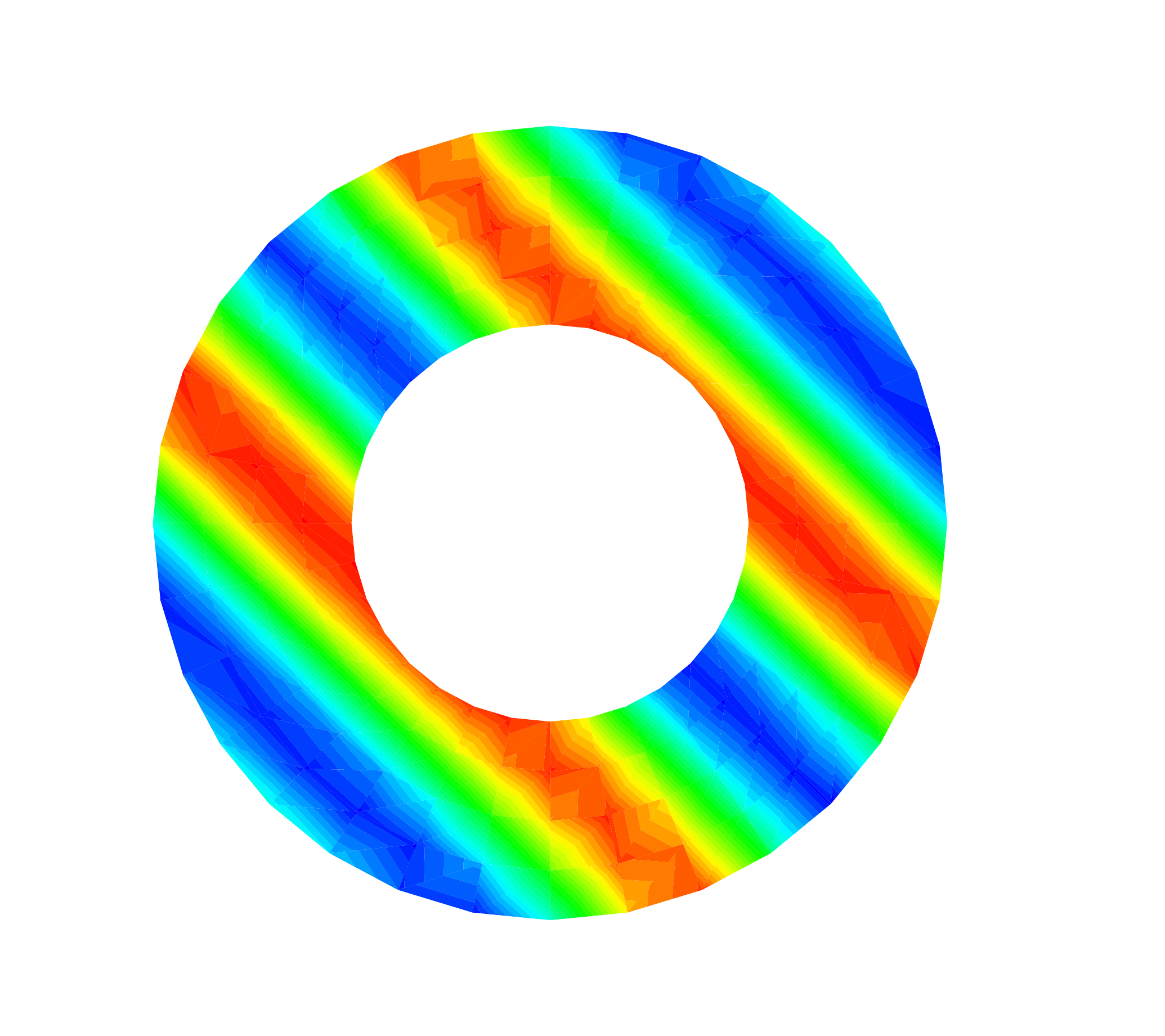}
    \includegraphics[width=0.32\textwidth, trim=20 20 40 50, clip]{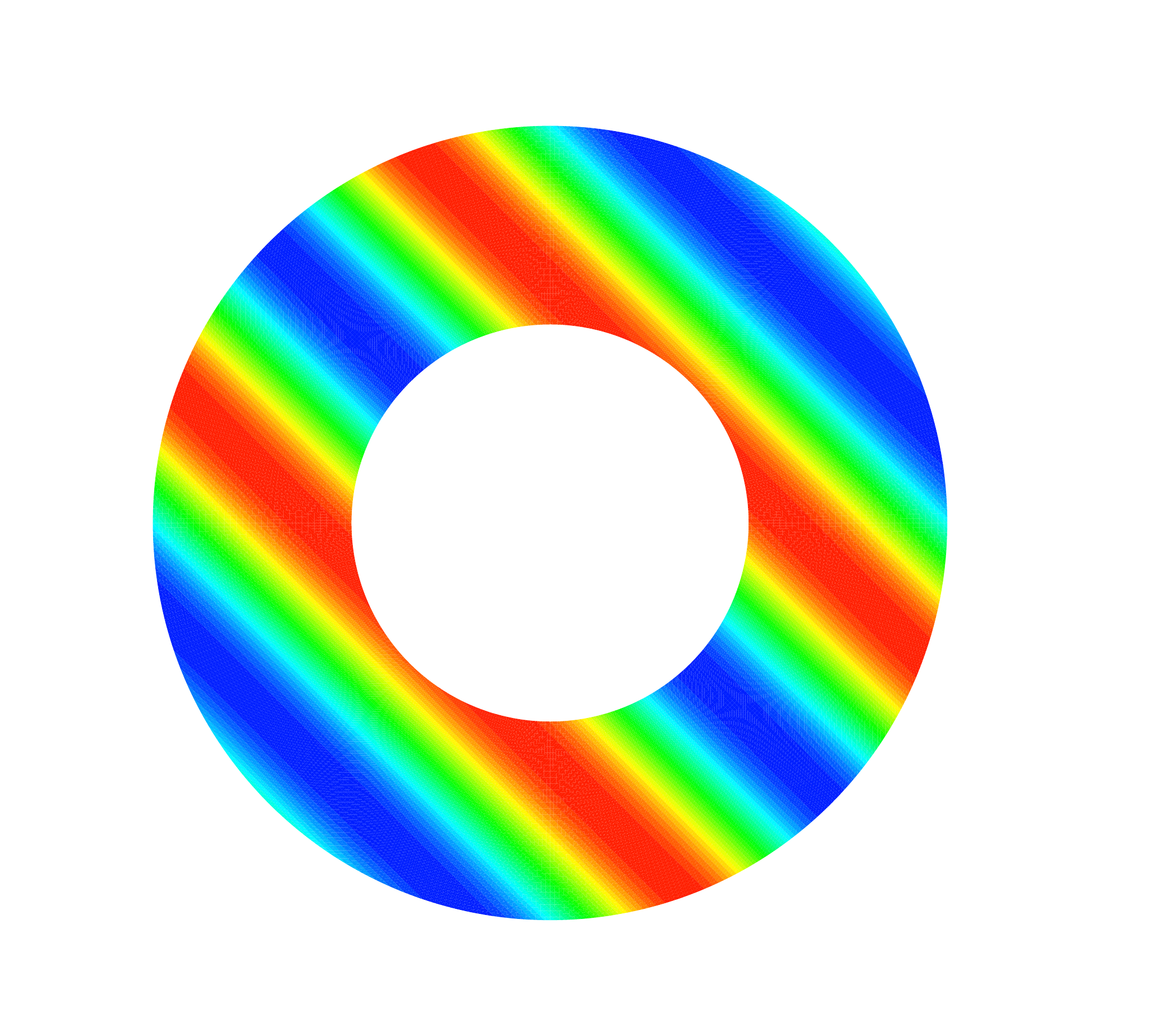}\\[2pt]
\includegraphics[width=0.8\textwidth, trim=0 10 0 480, clip]{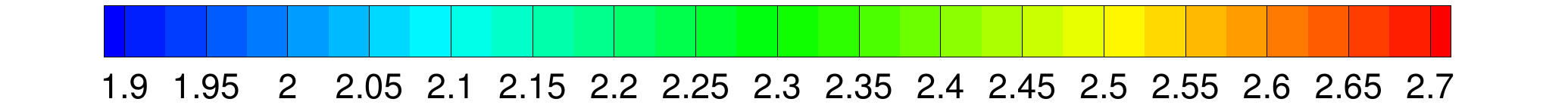}
  \caption{Circular domain and piecewise linear (k=1) DG discretization: Level-2 mesh and the initial condition (left), final solution on the level-2 mesh (middle), final solution on the level-6 mesh (right).}
  \label{fig:ring}
\end{figure}

The second test problem uses the same analytical solution as the previous one, but the domain is now ring-shaped with the inner radius of 500, outer radius of 1000, and centered at the origin. The ring is divided into 8 patches corresponding to MPI ranks. Fig.~\ref{fig:ring} shows the coarsest mesh with the initial condition (left), the final solution on the coarsest mesh (level-2, middle), and the final solution on the finest mesh (level-6, right) for the piecewise linear (k=1) DG discretization.

%\todo[inline]{domain size}
%8 patches / mpi ranks

\section{Conclusion and Outlook}

This work introduced for the first time quadrature-free functionality for a~SWE model based on the discontinuous Galerkin discretization.
In the future work, we'll demonstrate the performance of this methodology using simulations of real-life problems on complex 2D domains and test its scalability and computational performance on a~range of different hardware architectures.
Furthermore, we plan to transfer our approaches to 3D coastal ocean models such as implemented in UTBEST3D~\cite{DawsonAizinger2005,AizingerPDPN2013,ReuterAK2015}.

To complete the code generation pipeline, we plan to provide and incorporate the UFL (unified form language) similarly to approaches used in other code generation frameworks %Fenics and Firedrake 
as abstract representation on layer 1.
In addition to that, full support for block-structured grids should be available soon. 
We currently also work on improving node-level performance and scalability.

%Note for future 
%work that one could start from an UFL-like layer 1 formulation and also implement the steps to the quadrature-free form as algebraic transforms.

\section*{Acknowledgements}
The authors acknowledge financial support by the German Research Foundation (DFG) through grants AI 117/6-1, KO 4641/1-1, and GR 1107/3-1.

\bibliography{refs}

\begin{thebibliography}{10}

\bibitem{Aizinger2004}
V.~Aizinger.
\newblock {\em A Discontinuous {G}alerkin Method for Two- and Three-Dimensional
  Shallow-Water Equations}.
\newblock PhD thesis, University of Texas at Austin, 2004.

\bibitem{AizingerBF2018}
V.~Aizinger, L.~Bungert, and M.~Fried.
\newblock Comparison of two local discontinuous {G}alerkin formulations for the
  subjective surfaces problem.
\newblock {\em Computing and Visualization in Science}, 18(6):193--202, 2018.

\bibitem{aiz02}
V.~Aizinger and C.~Dawson.
\newblock A discontinuous {G}alerkin method for two-dimensional flow and
  transport in shallow water.
\newblock {\em Adv Water Resour.}, 25(1):67--84, 2002.

\bibitem{AizingerPDPN2013}
V.~Aizinger, J.~Proft, C.~Dawson, D.~Pothina, and S.~Negusse.
\newblock A three-dimensional discontinuous {G}alerkin model applied to the
  baroclinic simulation of {C}orpus {C}hristi {B}ay.
\newblock {\em Ocean Dynamics}, 63(1):89–113, 2013.

\bibitem{UFL}
M.~S. Aln{\ae}s, A.~Logg, K.~B. {\O}lgaard, M.~E. Rognes, and G.~N. Wells.
\newblock Unified form language: {A} domain-specific language for weak
  formulations of partial differential equations.
\newblock {\em ACM Trans. on Mathematical Software (TOMS)}, 40(2):9:1--9:37,
  2014.

\bibitem{asai2016coupled}
M.~Asai, Y.~Miyagawa, N.~Idris, A.~Muhari, and F.~Imamura.
\newblock Coupled tsunami simulations based on a 2d shallow-water
  equation-based finite difference method and 3d incompressible smoothed
  particle hydrodynamics.
\newblock {\em Journal of Earthquake and Tsunami}, 10(05), 2016.

\bibitem{Atkins1998}
H.L. Atkins and C.-W. Shu.
\newblock Quadrature-free implementation of discontinuous {G}alerkin method for
  hyperbolic equations.
\newblock {\em AIAA Journal}, 36(5):775--782, 1998.

\bibitem{bader2010dynamically}
M.~Bader, C.~B{\"o}ck, J.~Schwaiger, and C.~Vigh.
\newblock Dynamically adaptive simulations with minimal memory
  requirement—solving the shallow water equations using {S}ierpinski curves.
\newblock {\em SIAM Journal on Scientific Computing}, 32(1):212--228, 2010.

\bibitem{bankole2015semi}
A.~O. Bankole, A.~Iske, T.~Rung, and M.~Dumbser.
\newblock A semi-implicit {SPH} scheme for the two-dimensional shallow water
  equations.
\newblock In {\em Proceedings of the 10th International SPHERIC Workshop,
  Parma, Italy}, pages 252--258, 2015.

\bibitem{BungertAF2017}
L.~Bungert, V.~Aizinger, and M.~Fried.
\newblock A discontinuous {G}alerkin method for the subjective surfaces
  problem.
\newblock {\em Journal of Mathematical Imaging and Vision}, 58(1):147–161,
  2017.

\bibitem{casulli1990semi}
V.~Casulli.
\newblock Semi-implicit finite difference methods for the two-dimensional
  shallow water equations.
\newblock {\em Journal of Computational Physics}, 86(1):56--74, 1990.

\bibitem{CockburnShu1998}
B.~Cockburn and C.~Shu.
\newblock The local discontinuous {G}alerkin method for time-dependent
  convection--diffusion systems.
\newblock {\em SIAM Journal on Numerical Analysis}, 35(6):2440--2463, 1998.

\bibitem{DawsonAizinger2002a}
C.~Dawson and V.~Aizinger.
\newblock The local discontinuous {G}alerkin method for advection-diffusion
  equations arising in groundwater and surface water applications.
\newblock In J.~Chadam, A.~Cunningham, R.~E. Ewing, P.~Ortoleva, and M.~F.
  Wheeler, editors, {\em Resource Recovery, Confinement, and Remediation of
  Environmental Hazards}, page 231–245. Springer, 2002.

\bibitem{DawsonAizinger2005}
C.~Dawson and V.~Aizinger.
\newblock A discontinuous {G}alerkin method for three-dimensional shallow water
  equations.
\newblock {\em Journal of Scientific Computing}, 22(1-3):245–267, 2005.

\bibitem{Dubiner1991}
M.~Dubiner.
\newblock Spectral methods on triangles and other domains.
\newblock {\em Journal of Scientific Computing}, 6(4):345--390, Dec 1991.

\bibitem{GOFBS15}
T.~Gysi, C.~Osuna, O.~Fuhrer, M.~Bianco, and T.~C. Schulthess.
\newblock {STELLA}: {A} domain-specific tool for structured grid methods in
  weather and climate models.
\newblock In {\em Proceedings of International Conference for High Performance
  Computing, Networking, Storage and Analysis (SC)}, pages 41:1--41:12. ACM,
  nov 2015.

\bibitem{HajdukHAR2018}
H.~Hajduk, B.~R. Hodges, V.~Aizinger, and B.~Reuter.
\newblock Locally {F}iltered {T}ransport for computational efficiency in
  multi-component advection-reaction models.
\newblock {\em Environmental Modelling \& Software}, 102:185--198, 2018.

\bibitem{kuckuk2018whole}
S.~Kuckuk and H.~K{\"o}stler.
\newblock Whole program generation of massively parallel shallow water equation
  solvers.
\newblock In {\em 2018 IEEE International Conference on Cluster Computing
  (CLUSTER)}, pages 78--87. IEEE, 2018.

\bibitem{lai2016parallel}
W.~Lai and A.~A. Khan.
\newblock A parallel two-dimensional discontinuous {G}alerkin method for
  shallow-water flows using high-resolution unstructured meshes.
\newblock {\em Journal of Computing in Civil Engineering}, 31(3), 2016.

\bibitem{li2017generalized}
P.-W. Li and C.-M. Fan.
\newblock Generalized finite difference method for two-dimensional shallow
  water equations.
\newblock {\em Engineering Analysis with Boundary Elements}, 80:58--71, 2017.

\bibitem{Lockard1999}
D.~Lockard and H.~Atkins.
\newblock Efficient implementations of the quadrature-free discontinuous
  {G}alerkin method.
\newblock In {\em Proceeding of 14th AIAA CFD conference. AIAA}, pages
  526--536, 1999.

\bibitem{fenicsbook}
A.~Logg, K.-A. Mardal, and G.~N. Wells.
\newblock {\em Automated Solution of Differential Equations by the Finite
  Element Method}, volume~84 of {\em Lecture Notes in Computational Science and
  Engineering (LNCSE)}.
\newblock Springer, 2012.

\bibitem{poppl2016swe}
A.~P{\"o}ppl and M.~Bader.
\newblock Swe-x10: An actor-based and locally coordinated solver for the
  shallow water equations.
\newblock In {\em Proceedings of the 6th ACM SIGPLAN Workshop on X10}, pages
  30--31. ACM, 2016.

\bibitem{Pueschel2011}
M.~P{\"u}schel, F.~Franchetti, and Y.~Voronenko.
\newblock {\em Spiral}, volume~4, pages 1920--1933.
\newblock Springer, 2011.

\bibitem{Rathgeber2016}
F.~Rathgeber, D.~A. Ham, L.~Mitchell, M.~Lange, F.~Luporini, A.~T.~T. Mcrae,
  G.-T. Bercea, G.~R. Markall, and P.~H.~J. Kelly.
\newblock Firedrake: {A}utomating the finite element method by composing
  abstractions.
\newblock {\em ACM Trans. on Mathematical Software (TOMS)}, 43(3):24:1--24:27,
  2016.

\bibitem{ReuterAK2015}
B.~Reuter, V.~Aizinger, and H.~K{\"o}stler.
\newblock A multi-platform scaling study for an openmp parallelization of a
  discontinuous {G}alerkin ocean model.
\newblock {\em Computers and Fluids}, 117:325 -- 335, 2015.

\bibitem{SKHTL18procieee}
C.~Schmitt, S.~Kronawitter, F.~Hannig, J.~Teich, and C.~Lengauer.
\newblock Automating the development of high-performance multigrid solvers.
\newblock In {\em Proc. of the IEEE}, 2018.
\newblock Special Issue From High Level Specification to High Performance Code;
  to appear.

\bibitem{schmitt2014exaslang}
Christian Schmitt, S.~Kuckuk, F.~Hannig, H.~K{\"o}stler, and J.~Teich.
\newblock Exaslang: a domain-specific language for highly scalable multigrid
  solvers.
\newblock In {\em Domain-Specific Languages and High-Level Frameworks for High
  Performance Computing (WOLFHPC), 2014 Fourth International Workshop on},
  pages 42--51. IEEE, 2014.

\bibitem{unat2011mint}
D.~Unat, X.~Cai, and S.~B. Baden.
\newblock Mint: {R}ealizing {CUDA} performance in 3d stencil methods with
  annotated {C}.
\newblock In {\em Proceedings of International Conference on Supercomputing
  (ICS)}, pages 214--224. ACM, jun 2011.

\bibitem{vacondio2017non}
R.~Vacondio, A.~Dal~Pal{\`u}, A.~Ferrari, P.~Mignosa, F.~Aureli, and S.~Dazzi.
\newblock A non-uniform efficient grid type for {GPU}-parallel shallow water
  equations models.
\newblock {\em Environmental Modelling \& Software}, 88:119--137, 2017.

\bibitem{weinzierl2014block}
T.~Weinzierl, M.~Bader, K.~Unterweger, and R.~Wittmann.
\newblock Block fusion on dynamically adaptive spacetree grids for shallow
  water waves.
\newblock {\em Parallel Processing Letters}, 24(03), 2014.

\bibitem{wintermeyer2017entropy}
N.~Wintermeyer, A.~R. Winters, G.~J. Gassner, and D.~A. Kopriva.
\newblock An entropy stable nodal discontinuous {G}alerkin method for the two
  dimensional shallow water equations on unstructured curvilinear meshes with
  discontinuous bathymetry.
\newblock {\em Journal of Computational Physics}, 340:200--242, 2017.

\bibitem{wittmann2017high}
R.~Wittmann, H.-J. Bungartz, and P.~Neumann.
\newblock High performance shallow water kernels for parallel overland flow
  simulations based on {FullSWOF2D}.
\newblock {\em Computers \& Mathematics with Applications}, 74(1):110--125,
  2017.

\bibitem{Zint2018}
D.~Zint, R.~Grosso, V.~Aizinger, and H.~K\"ostler.
\newblock Generation of block structured grids on complex domains for high
  performance simulation (accepted).
\newblock In {\em Proceedings of the 9th International Conference on Numerical
  Geometry, Grid Generation and Scientific Computing ({NUMGRID2018})}, 2018.

\end{thebibliography}
\bibliographystyle{plain}
%\printbibliography

\end {document}